\documentclass[dvips]{arxstspdf}
\usepackage{multirow}
\usepackage{graphicx}
\usepackage{flushend}
\usepackage{stfloats}

\volume{24}
\issue{4}
\pubyear{2009}
\firstpage{472}
\lastpage{488}
\doi{10.1214/09-STS287}

\begin{document}
\begin{frontmatter}

\title{Structures and Assumptions: Strategies to~Harness Gene $\times
$ Gene and Gene $\times$ Environment Interactions in GWAS}
\runtitle{Interactions in GWAS}

\begin{aug}
\author[a]{\fnms{Charles} \snm{Kooperberg}\ead[label=e1]{clk@fhcrc.org}\corref{}},
\author[b]{\fnms{Michael} \snm{LeBlanc}\ead[label=e2]{mleblanc@fhcrc.org}},
\author[c]{\fnms{James Y.} \snm{Dai}\ead[label=e3]{jdai@scharp.org}} and
\author[d]{\fnms{Indika} \snm{Rajapakse}\ead[label=e4]{irajapak@fhcrc.org}}
\runauthor{Kooperberg, LeBlanc, Dai and Rajapakse}

\affiliation{Fred Hutchinson Cancer Research Center,
Fred Hutchinson Cancer Research Center,
Fred Hutchinson Cancer Research Center and
Fred Hutchinson Cancer Research Center}

\address[a]{Charles Kooperberg is Full Member, Division of Public Health Sciences,
Fred Hutchinson Cancer Research Center,
1100 Fairview Ave N / M3-A410, Seattle, WA 98109-1024 \printead{e1}.}
\address[b]{Michael LeBlanc is Full Member,
Division of Public Health Sciences,
Fred Hutchinson Cancer Research Center,
1100 Fairview Ave N / M3-C102,
Seattle, WA 98109-1024 \printead{e2}.}
\address[c]{James Y. Dai is Assistant Member,
Division of Public Health Sciences,
Fred Hutchinson Cancer Research Center,
1100 Fairview Ave~N / M2-C200,
Seattle, WA 98109-1024 \printead{e3}.}
\address[d]{Indika Rajapakse is Postdoctoral Fellow,
Division of Public Health Sciences,
Fred Hutchinson Cancer Research Center,
1100 Fairview Ave~N / M2-B500,
Seattle, WA 98109-1024 \printead{e4}.}
\end{aug}

%
\begin{abstract}
Genome-wide association studies, in
which as many as a million single nucleotide polymorphisms (SNP)
are measured on several thousand samples, are quickly becoming a common
type of study for identifying genetic
factors associated with many phenotypes.
There is a
strong assumption that interactions between SNPs or genes
and interactions between genes and environmental factors
substantially contribute to the
genetic risk of a disease. Identification of such interactions could
potentially lead to increased understanding about disease mechanisms; drug
$\times$ gene interactions could have profound applications for personalized
medicine; strong interaction effects could be beneficial for risk prediction
models. In this paper we provide an overview of different approaches to model
interactions, emphasizing approaches that make specific use of the
structure of genetic data, and those that make specific modeling
assumptions that may
(or may not) be reasonable to make. We conclude that to identify interactions
it is often necessary to do some selection of SNPs, for example, based
on prior hypothesis or marginal significance, but that to identify
SNPs that are marginally associated with a disease it may also be
useful to
consider larger numbers of interactions.
\end{abstract}

%

\end{frontmatter}

\section{Introduction}
Genome-wide association studies (GWAS), in
which as many as a million single nucleotide polymorphisms (SNP)
are measured on several thousand samples, are quickly becoming common
for identifying genetic
factors associated with many phenotypes. Until now most analyses of
GWAS have taken a one-SNP-at-a-time approach, some analyses are employing haplotypes,
but this is mostly as surrogates for unmeasured SNPs. There is,
however, the
strong assumption that interactions between SNPs or genes (gene $\times
$ gene
interactions) and interactions between genes and environmental factors
(gene $\times$ environment interactions) substantially contribute to the
genetic risk of a disease (e.g., Frankel and Stork, \citeyear{Frankel}; Philips, \citeyear{Philips}).
Identification of such interactions could
potentially lead to increased understanding about disease mechanisms; drug
$\times$ gene interactions could have profound applications for personalized
medicine; strong interaction effects could be beneficial for risk prediction
models.

The GWAS that have been carried out to date have
not lead to the identification of many such interactions, other
than that some of the SNPs, that were
found to be marginally associated with a disease, are
sometimes also tested for differences in subgroups (e.g., Gudbjartsson et al., \citeyear{Gudbjartsson}, Table 2). There are
several reasons for this lack of identification of interactions:
\begin{enumerate}
\item The odds ratios of SNPs that have main effects are usually small,
and there is no
reason to expect that interaction effects are any bigger. Increased
degrees of
freedom for models involving both main effects and interaction effects
result in
even lower power to specifically identify the interactions than the
already low power to identify
main effects.
\item The potential number of gene $\times$ gene
interactions is very large, for example, $\sim$\mbox{$5\times10^{11}$}
two-SNP interactions for the 1 million SNP
chip, making it computationally impossible to do
anything more than the simplest model for all interactions, and
reducing the
already limited
power further because of the required multiple comparisons correction.
\item The potential number of gene $\times$ environment interactions is
smaller; however, when there are several environmental factors of interest
the number of multiple comparisons for which we have
to correct is still considerably larger than the number of SNPs in the
initial scan. In
addition, some environmental factors may have measurement error, further
reducing power to identify interactions involving those factors.
\item With single SNP models there are a variety of (genetic) models to choose
from: for example, additive, recessive, dominant and codominant models.
For interaction models
there are many more, yielding an unwieldy array of models
and approaches to choose from,
some of which probably could be dismissed as not having the power to
identify interactions from the start. An example of such a model
with little power is one for $k$th order interactions, with $k>2$,
where all $3^k$
possible combinations of SNPs are modeled separately.
\item Imputation methods for marginal SNP effects work fairly well in
identifying
disease associated SNPs that are unmeasured, although the power for
these unmeasured SNPs is somewhat lower than for measured SNPs; for
interactions it becomes infeasible to check all
possible interactions involving unmeasured SNPs.
\end{enumerate}

%
\begin{figure*}[b]

\includegraphics{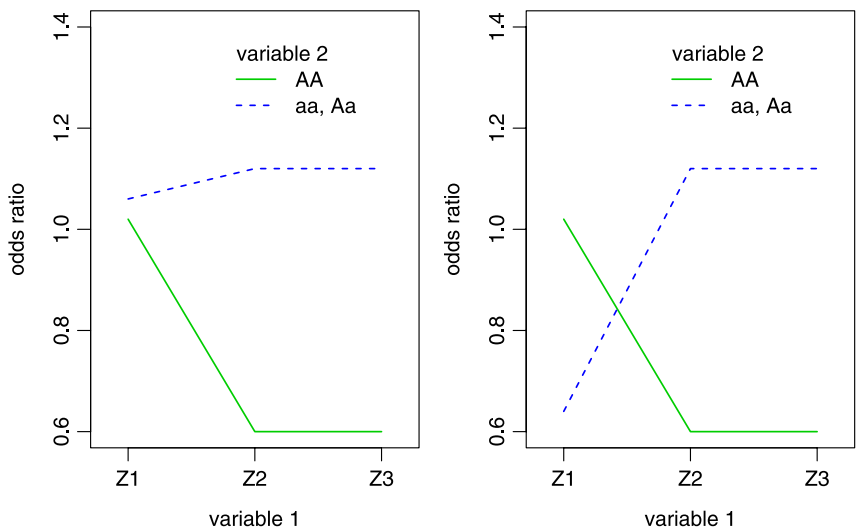}

\caption{Two different scenarios of interaction effects. In the left
panel disease risk is always higher
in the $\{\mbox{aa},\mbox{Aa}\}$ group; in the right panel the
disease risk actually is reversed for variable
level $Z1$ versus $Z2$ and $Z3$.}\label{intfig}
\end{figure*}

In this paper we will review a number of approaches that can be used
for finding
interactions in GWAS. We will use a few guiding principles when using
these methods:
\begin{itemize}
\item\textit{If a main effect fits the data well enough, don't use an interaction.}
This may seem obvious,
and many established function estimation methods adhere to this
principle. However, not
all models for genetic interactions adhere to this keep-it-simple principle.
\item\textit{Think about which model to use; think how the method scales up.}
This is another reason to keep-it-simple. If the model is too computer
intensive, it may not be feasible to fit it many millions of times.
\item\textit{If you need to divide the cake, give a slightly larger crumb to
the people who will enjoy it---that is, spend your power on the most
likely interactions.} Good candidates for SNPs
involved in gene $\times$ gene or gene $\times$ environment
interactions are prior hypothesized effects (if
there are any) and SNPs that show main effects. (Partly) ignoring other possible
interactions will eliminate the possibility of identifying these
ignored ones as significant;
since the power is very small to start out, we may as well use it wisely
and at least identify the more likely interactions.
\item\textit{Possibly insignificant interactions may help us to identify
disease associated genes.} If there is some difference in genetic
effects, looking
wisely in subgroups may help you find groups where the effect is strongest.
\item\textit{Be willing to exploit the genetic structure} [e.g.,
linkage (dis)equilibrium, SNPs taking only three values], \textit{be willing
to make some assumptions} (e.g., gene environment independence),
\textit{but be very aware what you loose if these assumptions are wrong.}
\end{itemize}
Obviously, many of these principles are not that different from studying
interactions in smaller problems: the main difference is that the size
of the problem for logistical reasons forces us to make the
right choice immediately---we may never get a second chance to correct
ourselves.

Furthermore, the nature of potential models of interactions has
implications both
for interpretation and statistical power. One class of methods uses models
leading to traditional odds ratio estimates of main effects and
interactions. These methods
are based on multiplicative interactions in a logistic regression and
can be represented
more generally as tensor product regression spline models. On the other
hand, another class
of strategies address the potential for strong associations within
subgroups of subjects;
these models include Logic Regression, tree-based regression, haplotype analysis
(we describe an adaptive regression strategy called SHARE in Section~\ref{sec3})
or adaptively weighted subgroup
analysis. We note that many methods which focus on subgroup effects
should not be interpreted
as definitively describing an interaction in the multiplicative sense
but rather as
tools to increase the potential of finding any association between gene
and outcome
or as tools for better risk prediction. We make the case, that the
choices of interaction
strategies which are potentially most powerful depend on the setting,
whether it is gene $\times$ gene
interaction studies, gene $\times$ environment studies, dimensionality
of the SNP or environmental data
and hypothesized genetic structure.

The next section starts by giving some general background about methods
for interaction
modeling in data analysis. In Section~\ref{sec3} we discuss some
methods for interactions that have
been developed for genetic studies.
In Section~\ref{twostage} we see how we can do some (limited)
identification of interactions
in GWAS, and in Section~\ref{sec5} we see how we can use interactions
in GWAS to identify
SNPs that are marginally associated with a disease.
We end with a brief discussion.

\section{Interactions in Statistical Modeling}\label{sec2}

Interactions are often characterized by departures from a simple additive
combination of effects in the context of some regression model. Such models
are of interest in a genetic association study, since one may like to describe
instances where a gene is associated with a disease only in the
presence of
another gene or in combination with an environmental factor. Alternatively,
modeling more complex models including interactions can improve function
approximations to derived better risk prediction models.

\subsection*{The ``Simplest'' Interaction Model}

To start the discussion,
consider a
simple interaction model involving two variables. A logit model for a binary
disease phenotype $Y\in\{0,1\}$ is
%
\begin{eqnarray}\label{simplest}
&&\operatorname{logit}[P(Y=1|\mathbf{X})]\nonumber
\\[-8pt]\\[-8pt]
&&\quad=\beta_{0}+\beta_{1}X_{1}+\beta_{2}X_{2}+\beta_{3} X_{1}X_{2},\nonumber
\end{eqnarray}
where $X_{1}$ and $X_{2}$ may code genes (SNPs or haplotypes). The
final term
in the model expresses a departure from a simple additive model, at
least on
the logit scale. In the setting of genomic association studies, interaction
models can also include two classes of variables, for instance, genetic and
environmental factors. In that case the model can be represented as
\[
\operatorname{logit}[P(Y=1|\mathbf{X,Z})]=\beta_{0}+\beta_{1}Z+\beta_{2}X_{1}+\beta
_{3}ZX_{1},
\]
where $Z$ denotes an environmental variable or some other subject exposure
variable, such as assigned treatment. The classic schema of a two variable
interaction is given in Figure~\ref{intfig} which shows the odds of
disease as a function of two variables. It shows two cases,
one where the effect of the second variable is only evident (or largely
evident) within a subset of levels of the first variable and a second
case, where
the effect is actually reversed at one level of the second variable. Sometimes
this second case is called a qualitative interaction, because unlike
the first
one, it does not depend on the choice of the link function to the outcome.
Again, the figures would apply to both gene $\times$ gene and gene
$\times$
environment interactions. Most would expect the effects shown in the
left panel to be the
most plausible
for gene by environmental effects in GWAS; that is, effect
modification, but not effect reversal.
A modest twist for GWAS is that one could
envisage 100s of thousands to millions of such interaction plots.

However, the simple model and Figure~\ref{intfig} hide important
aspects to
the complexity of actually constructing statistical interaction models in
several\break ways: (1)~the variables
that are involved in the interaction may only be observable as a derived
function of several variables; that is, in the models given above, the terms
$X_{i}$ or $Z$ may represent parametric or nonparametric functions of several
other variables, for instance, several SNPs representing a gene (2) the system
of predictors could be very high dimensional with respect to variable~$X$
(for instance, the many SNPs or genes in GWAS) and (3) even after selecting
variables, there are other problem specific issues like phasing,
measurement error and missingness. For point 1 and more recently for
point 2, there is extensive statistical literature addressing function
approximation and prediction modeling to draw upon and potentially use, at
least in concept, for the analysis of GWAS.

\subsection*{Additive Expansion Interaction Models}

The simple interaction model
above can be extended to a broader class of models for nonlinear,
nonadditive, multivariate regression methods. Assume that the disease model
indexed by regression function $\eta(\mathbf{X})$ is in some
$K$-dimensional linear space $\mathcal{B}(\mathcal{X})$,
%
\begin{equation}\label{expansion}
\eta(\mathbf{X})=\sum_{i=1}^{K}\beta_{i}g_{i}(\mathbf{X})
\end{equation}
for a given set of basis functions $g_{1}(\mathbf{X}),\ldots,g_{p}(\mathbf{X})$. Several
nonparametric multivariate regression\break methodologies use this additive
expansion (often\break called a basis function expansion). We review three common
function approximation methods that express the \mbox{basis} functions as tensor
products of individual covariates. While these methods should probably
not be directly applied to whole genome data for both computational and
statistical reasons, they can be useful after selecting smaller subsets of
variables. More importantly, they
follow the common and important paradigm, useful
for modeling interactions from GWAS, which involves searching for models
sequentially by \textit{first identifying main or (locally) marginal
effects before
fitting higher order terms.}
\begin{longlist}[\textit{(a)}]
\item[\textit{(a) Logistic Regression.}] Presented with more than two predictor
combinations, several variables may yield better predictions of the outcome
variable. Assuming modeling disease probability is a goal, the expansion
model (\ref{expansion}) would have component functions $g_{i}(\mathbf{X})$
which include products of two or more predictors, for example,
$g_{i}(\mathbf{X})=X_{j}X_{k}$.
An example model with at most 2-way interactions is
\[
\eta(\mathbf{X})=\beta_{0}+\sum_{j}^{p}\beta_{j}X_{j}+\sum_{j\leq
k}^p\beta
_{jk}X_{j}X_{k},
\]
where $\eta(\mathbf{X})$ may represent the conditional logit of the
probability of disease. Even with this simple model form,
the number of potential
interaction models is order $p^{2}$. If one does not limit the
interactions to only
involve pairs of variables, there are order $2^p$ models. The
numbers quickly rise with high order interactions and numbers of variables.
Therefore, even if we keep the model sparse with only a few $\beta
_{j}$ and
$\beta_{jk}$ not equal to zero, one would anticipate advantages both in
computation and variance control by constructing a reasonable pathway though
the model space. A standard approach is to use forward variable
selection and
consider adding interaction terms if one or more of the variables is already
included in the model as a main effect term. The above model is often not
sufficiently flexible for prediction modeling especially if $X_{j}$ are
multilevel or continuous. We consider some alternatives below.
\item[\textit{(b) Regression Spline Methods.}]
One possibility is that the $g_{i}(\mathbf{X})$ in the expansion
model (\ref{expansion}) are tensor products of piecewise linear
splines. Some
examples of methods using such an
expansion are Multivariate Adaptive Regression Splines
(MARS, Friedman, \citeyear{friedman1991})
and related spline methods Hazard Regression HARE and
PolyMARS (e.g., Kooperberg et al., \citeyear{CLK97}).

Regression spline algorithms exploit lower marginal or
additive structure to guide the search for interactions In addition, these
methods use basis functions that are tensor products of basis functions
in one
dimension. For example, if $g_{1}(\mathbf{X})=b_{1}(X_{k})$ and
$g_{2}(\mathbf{X})=b_{2}(X_{l})$ are two basis functions that depend
on a single
predictor, the $g_{3}(\mathbf{X})=b_{1}(X_{k})b_{2}(X_{l})$ is a tensor
product basis function that depends on two predictors. Truncated linear basis
splines are used to deal with continuous or ordered covariates
$(X_{i}-t_{k_{i}})_{+}=(X_{i}-t_{k_{i}})I\{X_{i}>t_{k_{i}}\}$.
Given that SNP data
only has only 3 categories, piecewise linear components are probably
most useful with respect to environmental factors in gene $\times$ environment
interaction models. Typically interactions of variables are included
only if
one or both of the variables are already identified as single variable terms
[e.g., $b_{1}(X_{k})$ or $b_{2}(X_{l})$] described above. This strategy yields
more interpretable models since the models contain main effects and it also
limits the search over the number of possible models, which better controls
variance compared to a search that evaluates all tensor products. The exact
restrictions on when tensor product basis functions are allowed in spline
models differs from one methodology to the other: for example, MARS
(Friedman, \citeyear{friedman1991})
has fewer restrictions than HARE (Kooperberg et al., \citeyear{Koop95}) and Polyclass and
PolyMARS (Kooperberg et al., \citeyear{CLK97}). All these methods identify lower order effects
first to control the search for higher order interaction terms and hence
control the variability of the search.

When building more complex models, one may also introduce
regularization or
penalization to reduce the impact of including too many parameters or cells
with very small counts of observations. Park and Hastie (\citeyear{ParkHastie2008}) develop an
algorithm that constructs penalized regression models for detecting gene
environmental interaction models which only includes interactions if one
of the main effects are in the model. Three level SNP variables are
coded as dummy variables and it uses quadratic penalization to stabilize the estimation.
\item[\textit{(c) Tree Regression.}] Regression trees are flexible methods
capable of
capturing interactions by recursively selecting and partitioning data
based on
low order associations. For tree based methods, the $g_{i}(\mathbf{X})$ in
the expansion model (\ref{expansion}) are indicator functions
corresponding to
rectangular regions, $R_{h}$, of the predictor space, $g_{h}(X)=I\{X\in
R_{h}\}$. The best known example in the statistical literature is
Classification and Regression Trees (CART, Breiman et al., \citeyear{cart84}).
Therefore, tree models
can be represented as a binary tree $T$, where the set of terminal nodes
$\widetilde{T}$ corresponds to the partition of the covariate space into
disjoint subsets. A~tree model can also be expressed by a basis function
representation
\[
\eta(\mathbf{X})=\sum_{h\in\widetilde{T}}\eta_{h}g_{h}(\mathbf{X}),
\]
where $g_h(\mathbf{X})$ is the region corresponding to a terminal node
$h$. This
function is a tensor product $g_{h}(X)=I\{X_{i}\in S_{1}\}\cdots\{
X_{p}\in
S_{p}\}$. To control the amount of computation, and to construct a limited
path through the large class of potential tensor product interaction
models of
this form, the model is grown in a forward stepwise fashion, similar to
stepwise regression. The assumption used by tree regression
is that effects can be found by searching
for \textit{a local marginal association} with outcome. The method is
applied to
the entire data set and predictor space, each variable and potential
split point
is evaluated. Of course if the predictors were just SNPs with coding $\{0,1,2\}$,
then only two splits are possible: one corresponding to a recessive
effect and one
corresponding to a dominant effect. The split point and variable that
leads to
the ``best'' split (as described below) is
chosen. The data and the predictor space are partitioned into two
groups. The
same algorithm is then recursively applied to each of the resulting groups.
Therefore, at any point on the regression tree, a split at a node $h$ yields
two nodes which can also be represented with the pair of basis functions
\begin{eqnarray*}
b_{hj}^{+}(\mathbf{X})
&=&I\bigl\{X_{h(j)}\in S_{h(j)}\bigr\}\quad\mbox{and}\quad
\\
b_{hj}^{-}(\mathbf{X})
&=&I\bigl\{X_{h(j)}\notin S_{h(j)},\bigr\},
\end{eqnarray*}
where $S_{h(j)}$ is a subset of the values of $X_{h(j)}$,
leading to terminal nodes basis functions $g_{h}(X)$ which products of such
indicator functions built up at each step. Typically a large tree is
grown to
avoid missing structure and then pruned back: model complexity is
reduced by
constructing a backward sequence of models using the cost-complexity pruning
algorithm.

\item[\textit{(d) Other Expansion Models.}] Of course, other
interaction outside the tensor product form of individual predictors
can be
expressed in this general form using parametric or nonparametric smooth based
component functions. For instance, multilayer neural networks construct the
$g_{i}(\mathbf{X})$ as composite function of functions of a linear
combinations of
subsets of the predictors $g_{i}(\mathbf{X})
=\phi(\sum_{i=1}^{p}\alpha_{i}X_{i})$. The regression tree methods
above can
also be extended to indicator functions based on linear combinations
$\{ \sum\alpha_{i}X_{i}>c\}$. One implementation, Flextree (Huang et al., \citeyear{Flextree}),
uses this model form.

It is clear that with more than a modest number of predictors the
potential number
of interaction models is huge and, hence, variance control is critical
in the
model search.
\end{longlist}

\subsection*{Bias/Variance Trade-Off: Picking Model Complexity}

Stepwise logistic regression, tree-based methods and adaptive regression splines
use a
forward selection strategy. A final model can be selected to minimize a
penalized measure of error,
\[
-l_{\alpha}=-l(\mathcal{M},\beta;Y_{i},\mathbf{X}_{i},i=1,\ldots
,n)+\alpha|\mathcal{M}|,
\]
where $l(\mathcal{M},\beta;Y_{i},\mathbf{X}_{i},i=1,\ldots,n)$ is
the fitted
log-likelihood for a model (of dimension $|\mathcal{M}|$) that was
considered, and $\alpha$ is a penalty parameter. Small penalty parameters
would lead to large models with limited bias, but potentially high variance;
larger penalty parameters lead to the selection of models biased toward the
null model, but with less variance. Given sufficient computation time,
a model
that minimizes the negative of the cross-validated likelihood (or some
resampling analog) may be the preferred method to address the bias versus
variance trade-off.

The above penalty only involves the number of parameters; often
other common penalties
can be useful. Additional terms that penalize either the $L^{1}$ norm of
the coefficients (e.g., LASSO, Tibshirani, \citeyear{Tibshirani96lasso}) or the
$L^{2}$ norm of the
coefficients (Ridge Regression, Hoerl and Kennard, \citeyear{HoerlKennard1970})
or a linear combination of the two
penalty terms (e.g., Elastic Net, Zou and Hastie, \citeyear{Zou05}) can lead to additional
effective ways to control variance.

While these are flexible statistical strategies for interaction model
building and
regularization, they do not directly incorporate any genomic structure
of the predictors.
As we describe in the next section, improved inferences, including
improved power for testing,
can be obtained by incorporating the special form of trinary SNP data
$0,1,2$, the nature of
dependence and/or independence between SNPs and environmental
variables, and haplotype structures.

\section{Models for Interactions in\break Genetic Studies}\label{sec3}

Genetic data have a number of special features that can be exploited in
modeling interactions.
In this section we discuss some
of those special features of genetic data, and how they have been used in
modeling interactions.
Most of these methods cannot deal with all SNPs as are commonly
measured in GWAS simultaneously. However, they can be directly used for
targeted regions
based on either prior biological hypothesis or top hits from an initial
single-SNP
filtering, as discussed in Section~\ref{twostage}.
It is hard to put the size restrictions of the various methods on one
scale. For example,
the SHARE method, discussed in Section~\ref{secLD}, is intended to find
interactions \textit{within} a block of SNPs in linkage disequilibrium,
and would thus be applied to
a fairly small number
of SNPs, for instance, 50 tag SNPs between
recombination hotspots. Nevertheless, for SHARE it is straightforward
to apply it to a GWAS
using a ``sliding window'' approach, where the method is applied to
overlapping blocks of SNPs
that
are close to each other in the genome.

On the other hand, methods like Multifactor
Dimensionality Reduction and Logic Regression, discussed in Section
\ref{secbinary},
are intended to find long-range interactions between a smaller number
of SNPs.
These methods
do not scale up to complete GWAS. However, they could be applied to
subsets of SNPs, like candidate gene studies, SNPs in a particular
pathway or SNPs that attain a
certain (marginal) significance level.
The number of SNPs that these methods can deal with is
typically up to a few hundred, or maybe a thousand, obviously depending
on the sample size
(so that the methods have enough power) and the available computing
resources (so that
sufficiently many models can be examined in a reasonable time). These
limitations are
quite understandable if we just consider the number of SNPs measured in
a GWAS. With, say,
one million SNPs, there are $5\times10^{11}$ two-SNP combinations. So,
even examining the
simplest interaction model for each combination of SNPs is expensive.

These size restrictions are much less severe for identifying gene
$\times$ environment interactions.
Again, this becomes clear from examining the scale of the problem.
Typically we will
only be interested in a few environmental factors, thus, the number of
potential single SNP
$\times$ environment interactions is smaller than the potential number
of SNP $\times$ SNP
interactions. Thus, some of the ideas on how to identify
gene $\times$ environment interactions discussed in
Section~\ref{indas} are directly applicable to GWAS.

To find gene $\times$ gene interactions in a GWAS, we have to take a
much simpler approach, for
example, the two-stage approach discussed in Section~\ref{twostage}.

\subsection{Genetic Data Is ``Almost Binary''}\label{secbinary}

Humans carry two copies of each chromosome, and
most genetic data comes from typing of Single Nucleotide Polymorphisms
(SNPs). SNP data is
commonly coded as 0/1/2, indicating the number of minor alleles at a
particular locus. If this
locus has a dominant effect on a disease phenotype, the genetic factor
$X$ can be coded 1 if
the SNP is 1 or 2, and 0 otherwise, and if this
locus has a recessive effect on a disease phenotype, the genetic factor
$X$ can be coded 1 if
the SNP is 2, and 0 otherwise. Dealing with binary data is attractive,
as models
are often easier to interpret, and many computations
can be done more efficiently.

This binary coding of genetic data is especially exploited
in Logic Regression (Ruczinski et al., \citeyear{LR}). The logic regression model is
\[
g[E(Y|\mathbf{X},\mathbf{Z})]=\beta_0 +\sum_{j=1}^m \beta_j L_j
+\sum_k \beta_{k+m}Z_k,
\]
where $Y$ is the disease response, $\mathbf{X}$ a vector of recoded
SNPs, $\mathbf{Z}$ a vector of other
(environmental) covariates, $g(\cdot)$ is a link function, such as the
logit, and the $L_j$ are
binary combinations of the $\mathbf{X}$, such as
\[
\bigl((X_1 \mbox{ and } X_7^c) \mbox{ or } X_3\bigr).
\]
The $L_j$ can be interpreted as risk factors. In Logic Regression model
selection is
carried out using permutation tests and cross-validation. In
particular, conditional
permutation tests are used to select simpler models when those fit the
data. An alternative
approach is to sample Logic Regression models using Markov chain Monte
Carlo (Kooperberg and Ruczinski, \citeyear{MCLR})
or bagging (Schwender and Ickstadt, \citeyear{BLR}).
The search among candidate models is carried out using a stochastic
simulated annealing
algorithm, though if the number of SNPs ($\mathbf{X}$) is limited and
the maximum number of SNPs
in each $L_j$ were limited to, say, 3, all models could be enumerated.

Logic Regression can be used to find gene $\times$ gene interactions,
and gene $\times$ environment
interactions for binary environmental variables. For example, in Kooperberg et al. (\citeyear{RAS}), for an
analysis of a candidate gene study of cardiovascular disease among
hypertensive, Logic Regression
was used to identify the model
\begin{eqnarray*}
&&\hspace*{-4pt}\operatorname{logit}[P(\mbox{myocardial infarction}|
\\
&&\qquad\hspace*{12pt}\mbox{AGTR2 SNPs, hypertensive drugs})]\\
&&\hspace*{-4pt}\quad=-0.90-0.72[(\geq1 \mbox{ A allele at rs171231429})
\\
&&\qquad\hspace*{60pt}\mbox{and }(\mbox{no calcium channel}
\\
&&\hspace*{174pt}\mbox{blockers})].
\end{eqnarray*}

As most methods described in this section, Logic Regression does not
scale up well to
the size of GWAS and needs
some selection of SNPs. The stochastic search algorithm could not
possibly examine a sufficiently large number
of models when there are hundreds of thousands of SNPs. Permutation
tests or cross-validation are even
more prohibitive.

While Logic Regression reduces SNPs from a 0/1/2 variable to a binary
variable, multifactor
dimensionality reduction (MDR, Ritchie et al., \citeyear{MDR}) makes use of the 0/1/2
nature of the
SNP data. For two specific SNPs MDR divides the nine
combinations into those that are associated with high and low risk for
a particular outcome. Thus, an MDR model
for two SNPs may be\vspace*{12pt}
%
%
\begin{center}
\begin{tabular}{@{}c c|c|c|c|@{}}
&\multicolumn{1}{c}{ }&\multicolumn{3}{c}{SNP A}\\[-1pt]
&\multicolumn{1}{c}{ }&\multicolumn{1}{c}{0}&\multicolumn{1}{c}{1}& \multicolumn{1}{c}{2}\\[-1pt] \cline{3-5}\noalign{\vspace*{-1pt}}
&0&H&H&L\\[-1pt]\cline{3-5}\noalign{\vspace*{-1pt}}
SNP B&1&L&H&H\\[-1pt]\cline{3-5}\noalign{\vspace*{-1pt}}
&2&L&H&L\\[-1pt] \cline{3-5}
\end{tabular}
\end{center}
where H and L refer to high and low risk, respectively. Three-level
interactions are modeled using
all 27 possible combinations of three SNPs, and so on. Among all
interactions up to a
certain level, MDR chooses the best combination by cross-validation. Dividing
the SNP combinations in high-risk and low-risk clearly has a close connection
to classification trees (see Section~\ref{sec2}), but the
nonmonotonicity for some SNP combinations
(e.g., for SNP B $=2$ in the example above) makes the method less
regularized and
maybe harder to interpret. The
cross-validation for MDR does not specifically prefer lower order
models over interactions.
Thus, for example, a two-level MDR may be chosen, by chance, if in fact
there are two main
effects. This leads to potentially increasing the type 1 error of
incorrectly identifying an
interaction (but not increasing the type 1 error of incorrectly
identifying a genetic effect).
Similarly, three SNP interactions may be identified when a model with
two SNP
interactions fits the model well. MDR has been applied
to a substantial number of candidate gene studies.

\subsection{Linkage Disequilibrium}\label{secLD}

 SNPs that are close to each other on the chromosome are
typically highly correlated, because of the shared ancestral history.
The extent of
this correlation or linkage disequilibrium (LD)
is known from databases such as the HapMap (HapMap Consotium, \citeyear{hapmap}) and the 1000 genomes
project (Kaiser, \citeyear{kaiser}) which is currently underway. Known LD structure
can be used to
impute SNPs that are not measured (e.g., Servin and Stephens, \citeyear{servin}; Marchini et al., \citeyear{marchini}).
LD can also be used to develop multilocus association
methods, often based on haplotype reconstructions (e.g., Browning and Browning, \citeyear{browning}; Epstein et al., \citeyear{epstein};
Lin and Zeng, \citeyear{lin}).
Statistically, the main effect of a haplotype can be deemed as a combination
of main effects and interactions in a locus (SNP) regression model
(Schaid, \citeyear{Schaid04}).
Using haplotypes can be an effective way to model interactions between multiple
mutations within a gene.
The latter perspective
may receive an increasing appreciation as genome-wide sequencing
technologies hold promise
to directly capture the rare variants in the near future. Several recent
candidate gene studies suggested that accumulation of multiple rare
alleles may contribute
to the risk of some common diseases (Vermeire et al., \citeyear{Vermeire2002}; Cohen et al., \citeyear{Cohen2004}; Nejentsev et al., \citeyear{Nejentsev}).
Theoretic population genetic studies also support the presence
of multiple deleterious susceptibility loci with low
frequencies (Pritchard, \citeyear{Pritchard2001}, Kryukov, Pennacchio and Sunyaev, \citeyear{Kryukov2007}).
To this end, a haplotype analysis can be useful
to assemble the cumulative, possibly interactive, effect of rare
variants within a gene.

Certainly locus-based models, such as the stepwise regression method
of Cordell and Clayton (\citeyear{Cordell02}), can be used to
model the local interactions directly without having to deal with
haplotype phase
ambiguity. However, with the high density of SNP data currently being
collected, haplotype
phase ambiguity is less a concern for power. Furthermore, the main
effect of a
haplotype already contains interactions when casted in locus
(SNP)-based models,
and hence tend to be more parsimonious than locus-based models that may require
high-order interaction terms. In situations where the LD is strong, additional
power gain can be achieved by fewer parameters used to characterize the genetic
risk profile. For instance, consider a situation with 2 SNPs in
complete LD and hence
3 haplotypes (00, 10, 11). A logistic regression with additive main effects
and interactions uses 3 parameters, whereas a logistic regression with
additive haplotype main effects uses one less parameter.
With more rare, recent variants (with less opportunity of recombination)
discovered through sequencing, it is anticipated that haplotype-based models
will be more cost-effective than locus-based models in modeling such
local interactions.

When many SNPs in a region (gene) are under investigation, the number
of haplotypes constructed from all SNPs can be excessively large.
A haplotype scan using a moving window of 5 to 10 SNPs may miss the interactions
between SNPs separated in the region, since the 3-D structure
of a protein can bring amino acids further apart into one functional domain.
Strategies to perform model selection seem inevitable in order to characterize
relevant genetic variants. Dai et al. (\citeyear{Dai09}) propose the {\sc Snp}-{H}aplotype
{A}daptive REgression ({SHARE}) algorithm that seeks the most
informative set of SNPs for genetic association in a targeted region by
growing and shrinking haplotypes with one more or less SNP in a
stepwise fashion.
Though it is not always ``optimal,'' the stepwise selection is a
rational choice
given the computational demand facing a large number of SNPs and the fact
that haplotypes we observe today were formed by sequential (stepwise)
mutations in
history. It is hard to imagine that there is no marginal effect for a
haplotype carrying disease risk. Contrary to the popular haplotype clustering
approaches (e.g., Seltman et al., \citeyear{Seltman01}; Durrant et al., \citeyear{Durrant04}; Browning and Browning, \citeyear{browning}), in the SHARE algorithm
both the trait and the genotypes guide the model selection process, and
the SNP
selection is irrespective of the order of the SNPs in the region
(gene).\
Cross-validation is used to select the best set of SNPs for a haplotype
analysis,
and phase ambiguity is accounted for by treating haplotype estimation
as a
part of the procedure.

\begin{figure}

\includegraphics{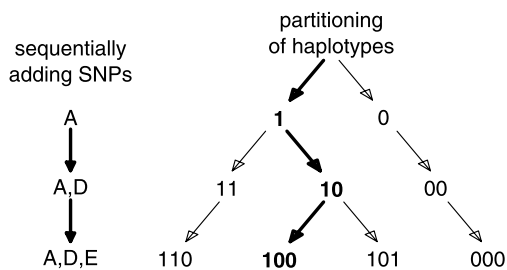}

\caption{Tree illustration of the sequential partition of haplotypes
when 5 SNPs $\mathbf{A}$--$\mathbf{E}$ are present. The left panel shows the growing set
of SNPs used
in the analysis and the right panel shows the partitions resulted from the
haplotypes based on the current set of SNPs. The minimal set of SNPs that
captures the genetic association is (A, D, E), with the disease risk
concentrated on the haplotype ``\texttt{100}''. The path leading to discovering
it is \texttt{1} $\rightarrow$ \texttt{10} $\rightarrow$ \texttt{100}. The
order of SNPs in the haplotypes follows $\mathbf{A}$, $\mathbf{AD}$ and $\mathbf{ADE}$ respectively.}\label{fig:share}
\end{figure}

Despite the resemblance to the stepwise
logistic regression of Cordell and Clayton (\citeyear{Cordell02}), SHARE is actually more closely
related to the Classification and Regression Tree (CART, Breiman et al., \citeyear{cart84})
algorithm (see Section~\ref{sec2}) in that it recursively partitions
the haplotype
sample space. Figure~\ref{fig:share} shows a simple example of the pathway
of partition. Among 5 SNPs in a region, we
first find the SNP \textbf{A} as the most significant SNP by a
single-locus scan,
so haplotypes are partitioned by ``0'' and ``1'' at the \textbf{A} locus.
Next, haplotypes constructed by \textbf{A} and \textbf{D} are found to be the
best 2-SNP haplotypes, and finally, the algorithm reaches the most informative
set \{\textbf{A}, \textbf{D}, \textbf{E}\} so that one 3-SNP haplotype concentrates the
disease risk.
To increase the chance of finding the most informative set, we grow longer
haplotypes and then prune back one SNP at a time. We use cross-validation
to select the partition with the minimal prediction error. While CART
is effective
to dissect high-order interactions, growing haplotypes is essentially
refining high-order interactions between loci. The difference between
SHARE and CART is that the recursive partitioning of CART is binary,
while SHARE potentially creates multiple splits when adding one SNP,
if there is recombination. Moreover, every subject has two sets of
haplotypes and we need some genetic model to describe the combined
haplotype effect, such as additive, dominant or recessive genetic
models.

\subsection{Independence Assumptions}\label{indas}

When it is known that pairs of SNPs or SNPs and environmental factors
are independent of
each other in the population, this knowledge can be effectively used to identify
interactions. Noticing that these independent quantities are not
independent in a
particular group of subjects (e.g., cases of a particular disease) now
implies an
interaction effect. This approach has been used to develop
methods for SNPs that are (assumed to
be) in linkage equilibrium, and for nongenetic environmental factors.

In randomized clinical trials
treatment
assignment is made independent of the genetic status of the participants.
In some other situations
it may also be reasonable to assume
gene $\times$ environment independence for some environmental
exposures. If for a case-control study this independence holds for the
controls (e.g.,
under the usual rare disease assumption), it is immediate that a
significant correlation
between a genetic effect and the environmental factor among cases
implies a
gene $\times$ environment interaction on the disease outcome. This is
the basic idea
behind the case-only analysis (Albert et al., \citeyear{Albert};
Piegorsch, Weinberg and Taylor, \citeyear{Piegorsch};
Umbach and Weinberg, \citeyear{Umbach}): there
is a simple relation between the odds-ratio among the cases and the parameter
for the interaction in a linear logistic regression model. Recently
a number of methods have been proposed to exploit this independence
also for
estimating main effects, and to avoid having to make the rare disease
assumption (Chatterjee and Carroll, \citeyear{Chatterjee}; Dai et al., \citeyear{Dai}).

It has been pointed out that violation of the
gene-environment independence assumption can seriously increase the
type 1 error
(Albert et al., \citeyear{Albert}). An empirical Bayes approach that ``averages'' the analysis
assuming independence and a traditional case-control analysis has been proposed
as a save alternative (Mukherjee and Chatterjee, \citeyear{Mukherjee}) that maintains some of the
advantage of the independence assumption, when that assumption cannot
be fully
confirmed. In this approach the effect estimate under the
(unbiased) case-control design is
averaged with the (more efficient) estimate using the case only
analysis, with
weights balancing the variance of the case-control estimator, and an empirical
Bayes estimate of the uncertainty of independence assumption.

Clearly, testing whether two factors are independent (e.g., gene and environment
among the controls) and then using a test with or without assuming
independence, depending
on the result of the independence test, will seriously inflate the type
1 error. On
the other hand, if a preliminary test is independent of the final test
for interaction
in the data set, such a test can be used to prioritize potential models
that are tested,
thereby alleviating the multiple comparisons problem. This is the
approach taken by
Millstein et al. (\citeyear{Millstein}). In this paper the authors first test for (gene-gene)
independence in the complete data set of cases and controls. Situations
where there
is substantial deviation of independence are prioritized for testing
for interaction
effects. The motivation seems to be that if two genes are dependent,
this dependence may
very well be different between cases and controls, for example, but not
necessary, because
the genes are independent among the controls but not among the cases.
As case-control
status is not used in this preliminary analysis, the eventual analysis
for interactions only
needs to be corrected for the interactions actually tested, which
increases power considerably.

\subsubsection*{Linkage equilibrium}

In a homogeneous population SNPs that
are on different chromosomes
or far apart on the same chromosome are (approximately) independent.
This suggests an alternative
way to identify interactions. For ``rare diseases,'' this
population-wise independence implies
that for controls the SNPs should be independent. Simple properties of
log-linear models show
that dependence between two SNPs among the cases now implies an
interaction effect of those SNPs on
case-control status. Zhao et al. (\citeyear{Zhao}) used this property to develop a
test for interactions by reconstructing haplotypes
(implicitly assuming Hardy--Weinberg equilibrium) between
two unlinked loci (SNPs) to get a measure of LD
between these SNPs. Assuming no LD among the controls, they
developed a (asymptotically $\chi^2$) test for an interaction effect
of these
two SNPs on a disease outcome.

In practice, it is unlikely that a population is completely
homogeneous. Thus, it
may be dangerous
to assume that SNPs are indeed in linkage equilibrium among the
controls. Given the large number
of SNPs that are typically tested, a small amount of correlation will
already inflate
the type 1 error rate. It is, however, a valid test of interaction in
genetic association studies
to test whether the correlation between two SNPs is the same among the
cases and the controls.
In fact, Zhao et al. (\citeyear{Zhao}) also provide a test for interaction using this
approach. However,
the examples in their paper that compare their approach to logistic
regression (which does not
use an independence assumption) make this
independence assumption.

\begin{table*}
\caption{Comparison of P-values for testing gene $\times$ gene interactions}\label{zhaotable}
\begin{tabular*}{\textwidth}{@{\extracolsep{\fill}}lccccc@{}}
\hline
&\multicolumn{3}{c}{\textbf{No independence assumed}}
&\multicolumn{2}{c@{}}{\textbf{Independence assumed}}\\[-6pt]
&\multicolumn{3}{c}{\hrulefill}
&\multicolumn{2}{c@{}}{\hrulefill}\\
& \textbf{Logistic}
& \textbf{Zhao et al.}
& \textbf{Rajapakse, Perlman}
& \textbf{Zhao et al.}
& \textbf{Rajapakse, Perlman}
\\
\textbf{Method}
& \textbf{regression}
& \textbf{(\citeyear{Zhao})}
& \textbf{and Kooperberg (\citeyear{Indika2009})}
& \textbf{(\citeyear{Zhao})}
& \textbf{and Kooperberg (\citeyear{Indika2009})}\\
\hline
TP53 $\times$ CD14& 0.0418&0.0325&0.0364& 0.1062 & 0.0210 \\
TNFR1 $\times$ APOC3&0.0001&0.0005&0.0005&0.0009&0.0000\\
TP53 $\times$ MDM2&0.0123&0.0715&0.0767&0.8607&0.0836\\
\hline
\end{tabular*}
\vspace*{-5pt}
\end{table*}

Rajapakse, Perlman and Kooperberg (\citeyear{Indika2009}) generalizes the approach by Zhao et al. (\citeyear{Zhao}) by constructing
a correlation matrix between groups of SNPs using the generalized
or composite LD of Weir (\citeyear{Weir}), separately for cases and
controls.
The advantage of using the generalized LD over other
measures of LD is that, since phase information is not required, no haplotype
reconstruction is needed, and
under certain conditions the generalized LD between two
SNPs reduces to the correlation between these SNPs when coded as 0/1/2.
Using this
approach, a simple test for identity between the correlation between
SNPs for
the cases and controls becomes a test of the interaction effect of
these SNPs
on case-control status.
There are a number of advantages for this approach. (i) This
test of independence can easily be extended to blocks of SNPs. An interaction
between two different blocks of SNPs on case-control status might
suggest that a haplotype, that may be a surrogate for an unmeasured
SNP in the first block, and a haplotype, that may be a surrogate for an unmeasured
SNP in the second block, have an interaction effect on case-control
status. Therefore,
this method is an alternative to the method of Chatterjee et al. (\citeyear{Tukey}) discussed
in Section~\ref{sec5}. (ii) A
test for identity between the complete correlation matrix for the cases
and the
controls is a global test for interactions among the SNPs considered on
case-control
status. (iii) Assumed independence between selected SNPs among the
controls can easily
be incorporated into this procedure by setting elements of the correlation
matrix for the controls equal to 0. When only two SNPs are examined, and
no independence assumption is made, we would expect the methods of
Zhao et al. (\citeyear{Zhao}) and Rajapakse, Perlman and Kooperberg (\citeyear{Indika2009}) to give similar
results, and that the results would be similar to the four
degree-of-freedom test
of comparing
\begin{eqnarray*}
&&\operatorname{logit}[P(Y=1|X_i,X_j)]
\\
&&\quad=\beta_0+
\beta_1 I(X_i=1)+
\beta_2 I(X_i=2)
\\
&&\qquad{}+
\beta_3 I(X_j=1)+
\beta_4 I(X_j=2)\\
&&\qquad{}+
\beta_5 I(X_i=1) I(X_j=1)
\\
&&\qquad{}+
\beta_6 I(X_i=2) I(X_j=2)\\
&&\qquad{}+
\beta_7 I(X_i=1) I(X_j=1)
\\
&&\qquad{}+
\beta_8 I(X_i=2) I(X_j=2)
\end{eqnarray*}
and
\begin{eqnarray*}
&&\operatorname{logit}[P(Y=1|X_i,X_j)]\\
&&\quad{}=\beta_0+
\beta_1 I(X_i=1)+
\beta_2 I(X_i=2)
\\
&&\qquad{}+\beta_3 I(X_j=1)+
\beta_4 I(X_j=2).
\end{eqnarray*}

In the implementation of Rajapakse, Perlman and Kooperberg (\citeyear{Indika2009})
these tests use the Kullback--Leibler
distance between two matrices. In Table~\ref{zhaotable} we provide
results on a previously
analyzed
case-control study
consisting of 779 heart disease patients, 342 of whom showed
restenosis, and 437 who did not (Hoh et al., \citeyear{Hoh2001}; Kooperberg and Ruczinski, \citeyear{MCLR}).
All individuals were genotyped for
89 SNPs/variants in 62 genes that were previously associated with heart
disease.
We show results for three of the two-SNP interactions that
were identified in Table III of Kooperberg and Ruczinski (\citeyear{MCLR}) (the other four
interactions in this table
involved a variant that had no homozygotic minor allele subjects). The
significance level for the
methods of Zhao et al. (\citeyear{Zhao}) and Rajapakse, Perlman and Kooperberg (\citeyear{Indika2009}) are based on 10,000
permutations. We note that,
as expected, the three methods give similar results when there is no
independence
assumption. When we do make an independence assumption, the approach of
Zhao et al. (\citeyear{Zhao})
appears less powerful than the one of Rajapakse, Perlman and Kooperberg (\citeyear{Indika2009}), which does
not require
a haplotype reconstruction. The approach of Rajapakse, Perlman and Kooperberg (\citeyear{Indika2009}) offers
the additional advantage
of potential for extension to tests of interaction effects between
blocks of SNPs.

\subsection{Using Main Effects to Find\break Interactions in GWAS}\label{twostage}

The methods discussed above exploit the
genetic structure of the data to be analyzed. However, mostly those
approaches do not scale up to GWAS, both because methods become
computationally too demanding, and because the number of
multiple comparisons becomes so large that the power
to identify significant interactions for anything other than
the strongest effects is missing. Interestingly, testing all models
that include an interaction for the combined effect of a particular
SNP on a disease outcome can increase
the power to identify an individual SNP as being associated
with a disease outcome (Marchini et al., \citeyear{Marchini2005}; Evans et al., \citeyear{Evans2006}), but it does
not increase the power to identify an interaction. We discuss this
approach further in Section~\ref{sec5}.

If only a few environmental factors are examined, the problem
to identify simple multiplicative gene $\times$ environment interactions
is essentially the same as studying marginal effects. Thus, while
power is limited, just like for any GWAS study, computationally
studying gene $\times$ environment interactions is straightforward.
However, the filtering procedures that we suggest below for gene
$\times$ gene
interactions can increase the power to identify
gene $\times$ environment interactions
in GWAS as well.

While it is probably obvious that enumerating all \mbox{interactions}
in a GWAS will be computationally too expensive, except for
the simplest possible models involving just
2-SNP interactions, adaptive search \mbox{algorithms}
do not circumvent these problems. Adaptive algorithms can be
roughly divided in those using stochastic search
algorithms and greedy search \mbox{algorithms}. Stochastic search algorithms,
like the simulated annealing algorithm used by Logic Regression, search
a stochastically selected set of models, selecting the best fitting
model(s) among those examined. For well structured model
classes these algorithms
can avoid looking at many poorly fitting models. However,
to have a reasonable opportunity to find good fitting models, the
number of models examined needs to increase considerably. In particular,
since in humans the extent of linkage disequilibrium is small
compared to the length of the genome ($r^2$ is typically much smaller than 0.8 after less than 50~kb;
Pritchard and Przeworski, \citeyear{pritchard2001}), the
number of models that are examined needs
to go up with close to the number of SNPs, and likely even more if
higher order interactions are studied to find good interactions. Greedy
search algorithms, for example, the stepwise and tree algorithms
described in Section~\ref{sec2}, are
much more likely to end up in local optimal models, that are
globally not very good,
when the number of predictors increases.

For these reasons, the most viable solutions to extend the methods discussed
above to GWAS would be to select SNPs based on some marginal criterion,
and only search for interactions among the selected SNPs. It is clear
that this
approach reduces the computational requirements. The two main questions are,
however, as follows:
\begin{itemize}
\item does a filtering procedure alleviate the multiple comparisons problem?
\item are we able to identify the ``important'' interactions?
\end{itemize}

As Marchini et al. (\citeyear{Marchini2005}) point out, the multiple
comparisons correction for testing interactions after marginal filtering
needs to take the filtering into account. The ``safe'' approach is to
correct the number of tests, for example, using a family-wide error
rate (FWER)
or a false discovery rate (FDR) approach for the number of interactions that
\textit{could} have been examined. Clearly, with such an approach the power
to identify significant interactions cannot be larger than when all
possible interactions \textit{would} have been examined
(but at reduced computational cost). This was part of what
was found by Marchini et al. (\citeyear{Marchini2005}) and Evans et al. (\citeyear{Evans2006}). We should
note here that, besides
that this is computationally infeasible, there is no simple
permutation test for (the strongest) interaction effect, as a simple permutation
of case-control status not only removes the interaction effect, but
also removes
all main effects. Such a permutation test would thus be a test of main effect
combined with interaction---a topic which we discuss in Section~\ref{sec5}.

\begin{table*}[b]
\tabcolsep=10.7pt
\caption{Sample power calculations for gene $\times$ environment interactions.
The numbers $0.00001,\ldots,1$ are the first stage significance $\alpha_1$}\label{tabpow}
\begin{tabular*}{\textwidth}{@{\extracolsep{\fill}}lccccccc@{}}
\hline
$\bolds{\beta_3}$ & \textbf{OR} & $\bolds{0.00001}$ & $\bolds{0.0001}$ & $\bolds{0.001}$ & $\bolds{0.01}$ & $\bolds{0.1}$ & $\bolds{1}$\\ \hline
\multicolumn{8}{@{}l}{Not assuming gene--environment independence}\\
0.262 & 1.3 & 0.21 & 0.26 & 0.18 & 0.10 & 0.04 & 0.02\\
0.336 & 1.4 & 0.66 & 0.72 & 0.56 & 0.37 & 0.21 & 0.11\\
0.405 & 1.5 & 0.94 & 0.94 & 0.85 & 0.69 & 0.51 & 0.34\\[6pt]
\multicolumn{8}{@{}l}{Assuming gene--environment independence}\\
\\
0.262 & 1.3 & 0.29 & 0.45 & 0.52 & 0.47 & 0.33 & 0.19\\
0.336 & 1.4 & 0.73 & 0.87 & 0.93 & 0.89 & 0.79 & 0.65\\
0.405 & 1.5 & 0.95 & 0.99 & 1.00 & 0.99 & 0.97 & 0.94\\
\hline
\end{tabular*}
\end{table*}

Kooperberg and LeBlanc (\citeyear{KLB2008}) establish that if the marginal testing of SNPs is carried
out using regression models of the form
\[
\gamma_0 + \gamma_1 X_i,
\]
for some coding $X_i$ of a SNP $i$ and the interaction model examined
is of
the form
%
\begin{equation}\label{ia.mod}
\beta_0 + \beta_1 X_i +\beta_2 X_j + \beta_3 X_iX_j,
\end{equation}
then the least squares estimates of $\gamma_1$ and $\beta_3$ are independent.
While the estimation in case-control studies is typically carried out using
logistic regression, this
independence result gives some justification of only correcting for
the number of tests that actually were examined, for
example, using a Bonferoni approach. Kooperberg and LeBlanc (\citeyear{KLB2008}) also
develop a resampling procedure based on scores, extending a technique
proposed by Lin (\citeyear{danyu}), that offers an alternative to permutation
tests, and is applicable to two stage studies in which only
SNPs that are marginally significant are tested for
interactions.
In this approach a sample from the efficient score for the interaction model
(\ref{ia.mod}) is generated
under the null hypothesis that $\beta_3=0$.

In particular, the efficient score for the addition of an interaction
$X_iX_j$, conditional on
$X_i$ and $X_j$ already being in the model, is $U_{ij}=\sum_k
U_{ijk}$, where
$U_{ijk}=(Y_k-p_{ijk})(X_{ik}X_{jk}-\mu_{ijk})$. Here
$Y_k$ is case-control status for subject $k$,
$p_{ijk}$ is the fitted probability of subject $k$ being a
case in a logistic
regression model using $X_i$ and $X_j$, but not $X_iX_j$, as predictor
for $Y$, and
$\mu_{ijk}$ is the fitted value for subject $k$ in the linear
regression model
using $X_i$ and $X_j$ as predictor for $X_iX_j$.
Under the null hypothesis of no association, $U_{ij}$ is approximately normal
with mean~0 and variance $V_{ij}=\sum_k U_{ijk}^2$. In computing the
significance level, we
compare $T=\max_{ij} U_{ij}^2/V_{ij}$ with
$T^*=\max_{ij}(\sum_k U_{ijk}Z_k)^2/V_{ij}$, where the $Z_k$ are
independent standard normal
random variables.
This approach does not assume
independence of the stage one and two tests, as the Bonferoni approach
does, but
rather the ``permutations'' are carried out conditional on the
results of the first stage.

A simple routine for
computing power calculations for two stage tests of interaction
is implemented in the CRAN package \texttt{powerGWASinteraction}.
Kooperberg and LeBlanc (\citeyear{KLB2008}) contain extensive simulation studies establishing that
the two-stage
approach indeed maintains the correct type 1 error rate, that the power in
most reasonable situations is vastly improved over a one-stage
analysis, and
that this power is well approximated by the routines from
\texttt{powerGWASinteraction}.

Clearly, this approach can also be applied to identify gene $\times$
environment
interactions: now only one SNP needs to be marginally significant to
be tested as part for a gene $\times$ environment interaction.
In Table~\ref{tabpow} we present power calculations for identifying a
gene $\times$ environment interaction. We assume model (\ref{ia.mod}) with
$X_i$ a binary SNP with $P(X_i=1)=0.4375$, which corresponds to a
dominant SNP
effect for a SNP with minor allele frequency 0.25, a binary environmental
factor $X_j$ with $P(X_j=1)=0.5$, a case-control study with 5000 cases
and controls,
500,000 SNPs, $\beta_0=-2$ (not a rare disease), $\beta_1=0$ (no
genetic effect when
$X_j=0$), $\beta_2=0.5$ (a moderate environmental effect), and an
overall multiple-comparisons
controlled significance level $\alpha=0.05$. We show results for
several gene $\times$
environment interaction effects $\beta_3$, and several levels for
the\break
marginal level of significance $\alpha_1$
that a SNP has to satisfy before it is tested for the gene $\times$
environment interaction.
Besides power to identify the
interaction using a regular analysis,
we also show power for an analysis that assumes
that the gene and the environmental factor are independent.

We see from Table~\ref{tabpow} that a two-stage procedure increases
the power
considerably. If only the top 500 SNPs are tested for gene $\times$
environment interactions
($\alpha_1=0.0001$),
the power to identify an interaction with odds ratio 1.4 is over 70\%. If
the gene and environmental factor are assumed to
be independent, and we analyze the data using,
for example, a case-only analysis or the approach of Dai et al. (\citeyear{Dai}), the
power increases
to about 90\%.

We applied the two-stage approach to the
WTCCC Crohn's disease data (WTCCC, \citeyear{Crohn}). We identified 211 SNPs that
had minor allele frequency $\geq0.1$, less than 5\% missing data and
marginal significance level $p<0.0001$, and did
not grossly violate Hardy Weinberg Equilibrium. Among the ${211\choose 2}=22155$ two-SNP interactions
among these
SNPs, three had a $q$-value${}< 0.05$ Storey and Tibshirani (FDR, \citeyear{Storey2003}),
and are thus plausible. Two of these interactions
involve SNPs on different chromosomes, the third one involves two SNPs
relatively close
together on the same chromosome.
Nineteen more possible interactions have $q$-values $< 0.25$, suggesting
that more
than ten of those show some reproducible association with Crohn's disease.

If the effect of a SNP goes in the opposite direction for two levels
of another SNP or an environmental variable,
a two-stage analysis may have less power than a one-stage
analysis (if these opposite effects just cancel each other out). We believe
that some of the more unusual interaction effects considered in Evans et al. (\citeyear{Evans2006})
are less likely, and we believe that using the power for more likely
scenarios is a potentially more fruitful way of $\alpha$-spending.

\section{Using Interactions to Find Main\break Effects in GWAS}\label{sec5}

Modeling gene $\times$ environment or gene $\times$ gene interactions
is useful even if the
goal of the analysis is primarily the identification of simple gene-disease
associations. For instance, if not acknowledged in the analysis method,
interactions can lead to attenuation of the marginal effect size and reduce
the power to detect true associations. Several authors have incorporated
interactions into their search of marginal genetic association. A common
thread of successful methods is that they allow model flexibility, but not
so much model flexibility as to substantially increase variance.

For instance, in the simple two variable model (\ref{simplest}) in
Section~\ref{sec2}, one can test the overall association of $X_{1}$
with disease outcome by
testing the null hypothesis $H_0:\beta_{1}=\beta_{3}=0$ using a 2
degree of freedom test.
Exploiting potential interactions to detect genetic association with this
simple testing technique has been explored by Kraft et al. (\citeyear{Kraft2007}). In a less
directed fashion, Marchini et al. (\citeyear{Marchini2005}) tested both the main effects
and interactions to explain the 3 $\times$ 3 table of two SNPs to
assess individual
associations.

Extensions to multiple predictors can substantially increase the
potential number of
parameters. For instance, consider two sets of variables $X_{1i}$ and $X_{2j}$,
which could represent two sets of SNPs or SNPs and environmental
variables. For assessing the association of a given $X_{1i}$ with outcome,
one can simultaneously test the $1+q$ terms $\beta_{1i}$ and
$\beta_{ij}$ (if there are multiple $X_{1i}, i=1,\ldots,p$,
testing involves $p+pq$ terms) in the model
\[
\eta(\mathbf{X})=\beta_{0}+\sum_{i=1}^{p}\beta_{1i}X_{1i}+\sum
_{j=1}^{q}\beta_{2j}X_{2j}+\sum_{i,j}\beta_{ij}X_{1i}X_{2j},
\]
where the components $X_{2j}$ could represent other SNPs or
environmental factors. The difficulty is that as $q$ increases, the
potential power of the test may significantly decrease due
to the increased number of parameters. One way to limit model
complexity is to specify a restricted form for the interaction model
such as
\begin{eqnarray*}
\eta(\mathbf{X})&=&\beta_{0}+\sum_{i=1}^{p}\beta_{1i}X_{1i}+
\sum_{j=1}^{q}\beta_{2j}X_{2j}
\\
&&{}+\theta\sum_{i,j}\beta_{1i}\beta
_{2j}X_{1i}X_{2j}.
\end{eqnarray*}
This class of models is used by Chatterjee et al. (\citeyear{Tukey}), and relates to the
idea of a one-degree of freedom interaction test, dating back to
Tukey (\citeyear{Tukey1949}). Suppose a gene is identified by $p$ SNPs $X_{1i}$,
then a
hypothesis of no association with outcome for this set $X_{1i}$ could be
phrased as $H_0{}\dvtx{}\beta_{1i}=0$, $i=1,\ldots,p$, where this indicates no
association through main effects or interactions since the $\beta_{1i}$
also appear in the interaction term. The strategy assesses overall variable
importance of a gene (potentially represented by several $X_{1i}$)
in the context of a more general model which could include interactions.
For instance, one could use any regularized and or stepwise model building
strategy (e.g., regression trees or regression splines described in
Section~\ref{sec2},
or ensembles of such models) and evaluate the impact of removing 
the gene of interest from the model. One technique to measure the
importance of the variable is to evaluate the difference in the fit or
log-likelihood compared to the fit with the gene permuted with respect
to all
other variables (e.g., Breiman, \citeyear{Breiman2001}). However, this last
strategy is likely
not computationally feasible in the context of GWAS.

An alternative idea is to modify simple gene association test statistics
by weighting them to take advantage of an interaction and increase
power of
the test statistic. For instance, one could focus on subgroups of
subjects to test for genetic association. In addition, if the procedure was
computational efficient, it could be an alternative to fitting full
interactions with maximum likelihood methods.

\begin{figure}

\includegraphics{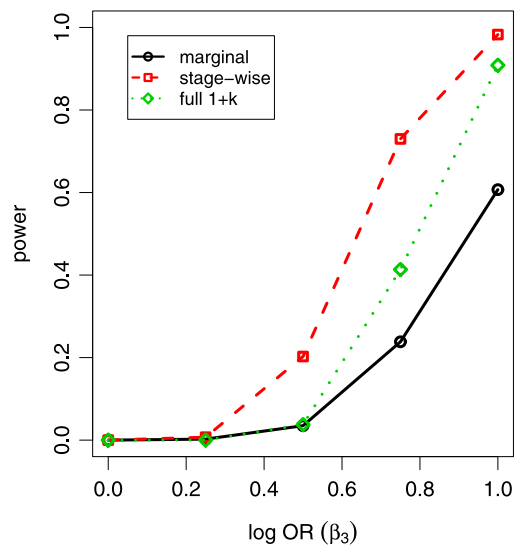}

\caption{Using interactions in tests of association: Power for
marginal test, full $1+k$ parameter association test, and stage-wise
weighted test with 2000 cases and 2000 controls and $\alpha=0.00001$.
The full $1+k$ parameter test
is based on testing the main effect and all interaction terms in a
logistic regression model
and the stage-wise test is based on using weighted score test
statistics where the
weights are derived from the Least Angle Regression (LARS) algorithm.
The main effect parameters in the data generating model are $\beta_{1}=0.05$ and $\beta_{2}=0$. }\label{powfig}
\end{figure}

Since many useful association tests are score type statistics, one can outline
the method in terms of weighted score test statistics. Let $Z$ denote an
environment or treatment variable and $X_{j}$ a genetic factor. For
example, in the case of binary outcome data, let $X_{ji}$ be the gene $j$
value and $Z_{ki}$ environmental factor $k$ for individual $i$, the
score component would be
$U_{ji}=X_{ji}(Y_{i}-\exp(\alpha+\beta Z_{ki}))/(1+(\alpha+\beta Z_{ki}))$.
If the association is thought to be stronger in a subgroup of subjects (e.g.,
heavier smokers) based on some environmental factors, then a subgroup weighted
marginal test statistic
\[
U_{W}(Z,X_{j};\theta)=\sum_{i=1}^{n}h(Z_{i},\theta)U_{ji},
\]
where $h(Z;\theta)=I\{Z>\theta\}$ and Z is an ordered environmental
variable, may have more power. The generalization to multiple or
alternative basis functions\looseness=1
\[
U_{W}(Z,X_{j};\theta)=\sum_{i=1}^{n}\sum_{k=1}^{q}\alpha
_{k}h(Z_{ik},\theta_{k})U_{ji}
\]
allows more flexibility. The basis functions, $h(Z_{ik},\theta_{k})$,
could be simple subset functions,
$h(Z_{ik};\theta)=I\{ Z_{ik}>c_{k}\}$, or piecewise
linear functions, $h(Z_{ik};\theta)=\{ Z_{ik}-c_{k}\}
^{+}$, similar to
those used in tree-based models and regression spline models described in
Section~\ref{sec2}. As the direction of association is usually
unknown, the weights $\alpha
_{k}$ can be derived from the data. LeBlanc and Kooperberg (\citeyear{LBK2008}) obtain weights using
stage-wise regression based on the Least Angle Regression algorithm
(LARS, Efron et al., \citeyear{Efron04leastangle}), where the score components are the
outcome variable.
The intent is to focus the test statistic on the environmental
combination which leads to maximal genetic association. They show that
weighting the association test statistics can
significantly increase power in many simulated situations where gene
$\times$ environment
interactions exist. As an example, Figure~\ref{powfig} shows simulations
for 2000 cases and 2000 controls generated from the logistic
interaction regression\break model
\eject
\[
\eta(Z,X) =\beta_{0}+\beta_{1}X+\beta_{2}h(Z)+\beta_{3}Xh(Z),
\]
where we assume a binary SNP, $X$ with frequency equal to 0.2, and
$h(Z)$ a
function of several environmental variables. There were five environmental
variables corresponding to the basis functions
$(1,\{Z_{j}<c_{0.25}\},\{Z_{j}<c_{0.50} \},\{Z_{j}<c_{0.75}\})j=1,\ldots,5$,
available for modeling and $h(Z)$ depended on the linear combination of
two of those variables:
$h(Z)=\{Z_{1}<c_{0.50}\}-\{Z_{2} <c_{0.50}\}+0.25$. We evaluated a
marginal test of the
$1+k$, $k=3 \times5$ parameters (including main gene effect and the
modifying variables), and a regularized stage-wise test.
The type I error was controlled to approximately 0.00001. The results
are presented
in Figure~\ref{powfig}. The parameter values main effects are $\beta
_{1}=0.05$ and $\beta_{2}=0$
so that the small genetic effect increases as $h(Z)$ and $\beta_{3}$
increase. This
would be a plausible scenario for effects in a GWAS, a genetic effect
that is more apparent within
a subgroup of subjects exposed to a set of environmental conditions.
This model is similar to
the hypothetical effects shown in the left panel of Figure~\ref{intfig}.

As the magnitude of the interaction effect increases,
using the more complex model and joint association testing
substantially increases
power over marginal testing. The full model weighting, depending on 16
basis functions and parameters,
suffers somewhat from increased variance. However, the stage-wise test
statistic performs the best of the
three methods by controlling the overall variance. The tuning parameter
for the stage-wise method was set to correspond to approximately 2.5
degrees of freedom.

Therefore, if there are one or a small number of well characterized
environmental factors
that are substantially modifying the association of a gene with disease,
statistical strategies which incorporate interactions, and jointly test
the main effect and the interaction, are useful for improving power
over\break marginal association tests.

While the adaptive weighting strategy is more computationally demanding
than calculating traditional score test statistics, it is feasible to
conduct the analysis on the
GWAS because each SNP calculation is independent and the Least Angle
Regression algorithm using a small number
of environmental variables is very efficient, if the tuning parameter
is set a priori as we suggest.
However, the impact on variance is potentially a greater concern, there
is still likely be some
power advantage of filtering on main effects to, say, a small number of
100--1000s, before applying
the adaptive method analogous strategy described in Section~\ref{sec3}.

\section{Discussion}
Identifying interactions is typically not the main goal of a GWAS analysis.
Interaction effects may teach us things
about the biology behind a disease, or they may be beneficial in
constructing predication models. However, at least
as important, interaction effects or differences in the gene (SNP) effect
between different subgroups may actually help us in identifying
the significant SNPs. Therefore, we believe that it is important to
pay some attention to the identification of interactions.

Because of the number of SNPs under consideration in a typical GWAS, it
is virtually impossible to identify gene $\times$ gene
interaction effects, unless additional assumptions
are being made. We believe that the most fruitful approach
is to first identify SNPs that are (marginally) associated
with a disease, and then examine interactions involving those SNPs.
Not only does this seem reasonable because SNPs that have an
interaction effect
typically will also show some modest main effect, it also adheres
to a basic premise in statistical modeling which reduces the variance
in model building: don't model
interactions without main effects.

After such initial filtering, there is a substantial number of
approaches that
can be used to identify interactions that make use of the specific
form of genetic data. This is a reasonable two-stage approach if the
methods to identify interactions are used to independent data than
what was used to identify the marginal significant SNPs.
If the same data is used, however, care has to be taken that the initial
selection of SNPs does not bias the inference about the interactions.
We have shown
that for a simple model this is possible---but this is certainly not generally
true.

The story for gene $\times$ environment interactions is similar. The problem
of identifying such interactions is ``smaller,'' but it is still so large
that some filtering will often increase the power.

Exploiting the genetic structures and
making additional assumptions, like gene $\times$ gene independence
among genes
on different chromosomes, among controls, or gene $\times$ environment
independence, can substantially increase the power to identify interactions.
Clearly, however, if the assumptions are not true, making those
assumptions can substantially increase the type 1 error. Thus, if
those assumptions are uncertain, an empirical Bayes approach like the
one by Mukherjee and Chatterjee (\citeyear{Mukherjee}) (see Section~\ref{sec3}) may be safer.

\section*{Acknowledgments}
The authors were supported in part by Grants\break
NIH R01-CA74841, NIH P01-CA53996, NIH U01-\break CA125489, NIH R01-CA90998 and
NIH T32-\break CA80416.

\vspace*{-1.5pt}
\end{document}